# Influence of the Preparation Process on Microstructure, Critical Current Density and $T_c$ of MgB$_2$ Powder-In-Tube Wires

Sonja I. Schlachter, Wilfried Goldacker, Johann Reiner, Silke Zimmer, Bing Liu, and Bernhard Obst*Abstract*— MgB$_2$ wires and tapes suffer from a limitation of the transport current densities, especially at low field, due to an unsufficient thermal stabilization caused by hot spots from a non-homogeneous microstructure, the presence of secondary phases and voids. For the preparation of MgB$_2$ wires and tapes two different precursor routes, using MgB$_2$ (ex-situ process) or Mg + B (in-situ process) are applied. The in-situ approach allows variation of many parameters in the wire preparation and the improvement of this approach is the central topic of this paper. We present an improved in-situ approach, the characterization of the filament microstructure, effects from stoichiometry and the corresponding superconducting properties. The transport current capability of such in-situ wires could be improved to the level of the reference ex-situ wires.

*Index Terms*— Borides, Critical Current, Microstructure, Wires and Tapes## I. Introduction

SINCE the discovery of superconductivity in magnesium diboride [1], [2] at temperatures far above the transition temperatures of the technical low temperature superconductors, many efforts have been made to produce high current carrying conductors in wire and tape geometry [3]. As MgB$_2$ is brittle and not ductile deformable, wires and tapes for technical applications have to be prepared as coated conductors (thin films) or with techniques like the powder-in-tube (PIT) technique. PIT MgB$_2$ wires and tapes with different sheath materials achieve high transport critical current densities as high as or even above $10^5$ A/cm$^2$ in self field at 4.2 K [4]-[8]. For PIT wires and tapes either prereacted MgB$_2$ powder or mixtures of elemental Mg and B powders are used as precursor, the two different routes are called ex-situ and in-situ, respectively. The ex-situ route offers an easy way of wire and tape processing, as the conductors need not neccessarily be heat treated to achieve high critical current densities as shown by Grasso *et al.* [6].

PIT MgB$_2$ wires and tapes have been prepared with different sheath materials like Fe, Ni, Cu, Ag, Nb, Ta, stainless steel or composites of different materials. However, the filaments often tend to the formation of hot spots due to inhomogeneities, voids (pores) or impurity phases. Some of the sheath materials like Cu and Ni are suited well for thermal stabilization of the conductors, however, due to chemical reactions with the filament not for application of a heat treatment.

One major challenge for wire and tape processing is the increase of the irreversibility field by addition of pinning centers or by texturing. Currently, the critical fields in bulk samples are much below the critical fields of thin films. Measurements indicate an upper critical field $H_{c2}(0)$ of thin films around 40 T and an irreversibility field $H^*(0)$ above 20 T [9]. In contrast to ex-situ preparation methods using commercial MgB$_2$ powder, in-situ approaches using elemental Mg and B powder mixtures allow variation of the precursors, i.e. control of impurities like oxygen, doping of ternary elements and variation of stoichiometry. The main goal of this paper was to improve the current carrying capability of in-situ wires to the level of the ex-situ wires and to investigate the influence of the preparation process on microstructure and the superconducting properties.

## II. Experimental

MgB$_2$/Fe wires were prepared with two different precursor routes. The two routes, ex-situ and in-situ, correspond to the formation of MgB$_2$ before and during wire preparation, respectivly. Ex-situ wires are prepared using commercial prereacted MgB$_2$ powder (Alfa Aesar) while elemental Mg (-325 mesh, 99.8 % purity) and B (crystalline, -325 mesh, 99.7 % purity) powder mixtures are used as in-situ precursors.

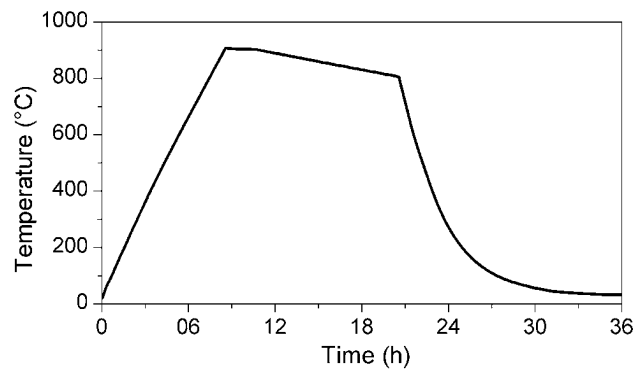

Fig. 1. Temperature profile of the MgB$_2$ formation heat treatment B.

Manuscript received August 6, 2002.
S. I. Schlachter is with the Forschungszentrum Karlsruhe, ITP, P.O. Box 3640, 76021 Karlsruhe, Germany (phone: +49-7247-825902; fax: +49-7247-825398; e-mail: sonja.schlachter@itp.fzk.de).
W. Goldacker, J. Reiner, S. Zimmer, B. Liu, and B. Obst are with the Forschungszentrum Karlsruhe, ITP, P.O. Box 3640, 76021 Karlsruhe, Germany.1

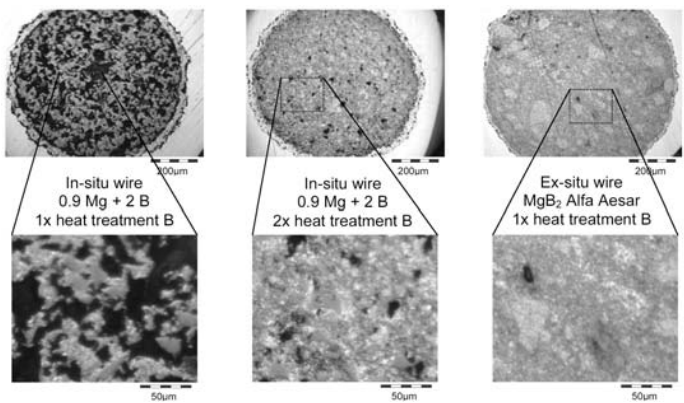

Fig. 2. Optical images of two in-situ and one ex-situ MgB$_2$/Fe wire.

Fe tubes (⌀=5 mm, wall thickness 0.5 mm) filled with the precursor were swaged to an outer diameter of 2.2 mm. At this diameter some of the wires were annealed for one hour at 600°C in vacuum (heat treatment A) while in the improved in-situ approach a first MgB$_2$ formation heat treatment (heat treatment B) in Ar(95%)/H$_2$(5%) atmosphere with the temperature profile shown in Fig. 1 was applied to the other wires. In a second deformation process all wires were swaged to the final diameter of approximately 1.1 mm and then heat treated with the temperature profile from Fig. 1.

As the melting point of Mg ($T_{m,Mg}$ = 650°C) is much below the temperature at which the final heat treatment for the MgB$_2$ formation has to be performed for in-situ wires, we varied the Mg : B ratio in order to compensate possible Mg losses due to evaporation or MgO formation during the heat treatment. The Mg : B ratio was changed from 0.7 : 2 up to 1.3 : 2.

The microstructure, especially the element distribution within the superconducting filaments was investigated by means of high resolution SEM/EDX. Short wire pieces (length 1 cm) were used to determine $T_c$ via ac-susceptibility measurements. The transport critical current density $J_c$ of short samples (length 4 cm) was measured with a standard 4-probe method at 4.2 K in magnetic fields up to 10 T, using a criterion of 1 μV/cm.

## III. RESULTS AND DISCUSSION

### A. Microstructure

Fig. 2 shows the optical images of the cross-sections of one ex-situ and two in-situ MgB$_2$/Fe wires (Mg : B = 0.9 : 2). The in-situ wire on the left and the ex-situ wire on the right were prepared with only one identical heat treatment (see Fig. 1), while for the improved process two such heat treatments were applied to the in-situ wire displayed in the middle part of Fig. 2. The filament cross-section of the in-situ wire on the left shows many big holes (black), while the filament of the improved in-situ wire and the filament of the ex-situ wire is rather dense. As the molar volume of MgB$_2$ ($V_{MgB2}$) is smaller than the sum of the molar volumes of the precursors ($V_{Mg}$ + 2·$V_B$) the filament of the in-situ wires shrinks during the first MgB$_2$ formation heat treatment. The further deformation to the final wire diameter by swaging removes occurring voids leading to dense filaments. No measurable further shrinking of the filament was observed after the second heat treatment of the wires. This volume effect is responsible for the poor filament densification using the former standard process with only one (final) reaction heat treatment.

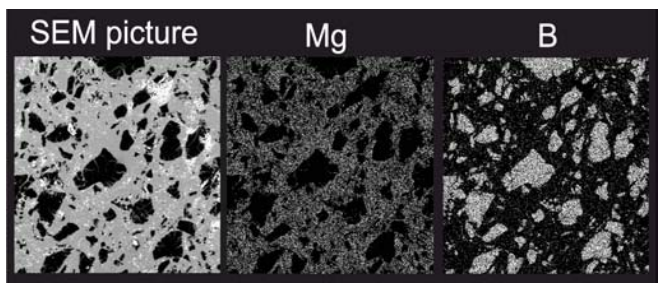

Fig. 3. SEM picture of an in-situ MgB$_2$/Fe wire with a precursor ratio of Mg : B = 0.9 : 2 (left) and EDX mapping showing the distribution of Mg (middle) and B (right) within the filament in white and grey color.

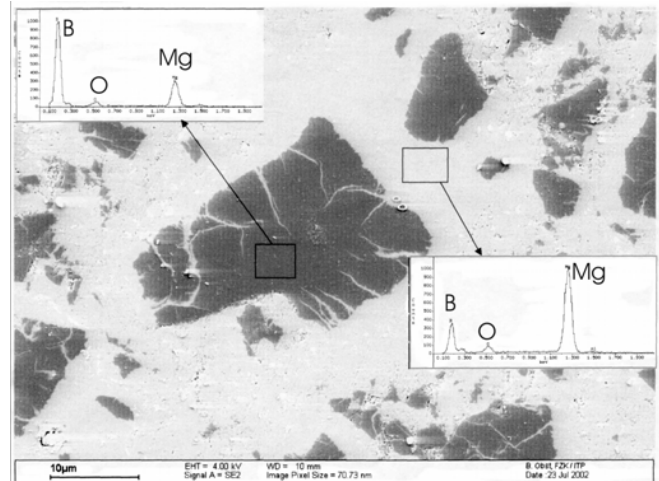

Fig. 4. SEM picture of an in-situ MgB$_2$/Fe wire with a precursor ratio of Mg : B = 0.9 : 2 showing boron rich grains (dark) in a matrix which is supposed to contain the major part of the MgB$_2$ phase. The insets show EDX spectra of the marked regions.

The disadvantage of the in-situ approach are reaction layers at the filament-sheath interface (Fig. 2).

Figs. 3 and 5 show SEM pictures of two wires from Fig. 2 and the distribution of Mg an B in the filament deduced from EDX mapping. The filament of the in-situ wire is very inhomogeneous and shows large boron rich grains up to approximately 30 μm length. EDX analysis of one of these grains (shown in Fig. 4) revealed that the marked area of this grain contains mainly boron (87.91 at%), little Mg (8.88 at%) and little oxygen (3.21 at%). However, Mg difuses into the grains along tiny cracks pathing through the grains like dendrites. Quantitative EDX-analysis shows that the area around the boron rich grains with 33.85 at% Mg, 5.08 at% O and 61.08 at% B could be the superconducting $MgB_2$ phase plus impurity phases like MgO.

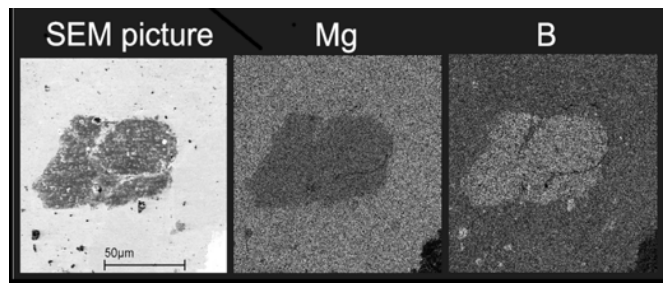

Fig. 5. SEM picture of an ex-situ MgB$_2$/Fe wire (left) and EDX mapping showing the distribution of Mg (middle) and B (right) within the filament in white and grey color.

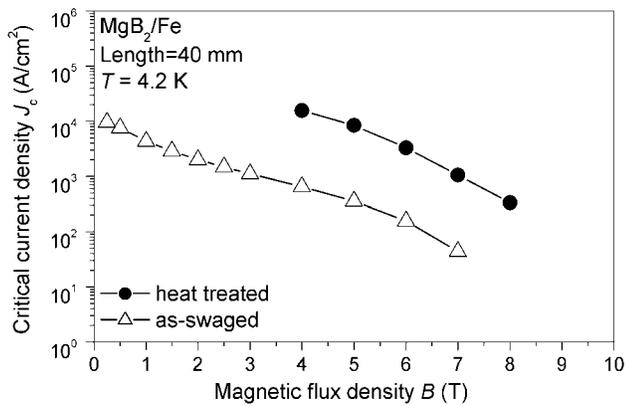

Fig. 6. Critical current density of as-swaged and heat treated ex-situ $MgB_2$/Fe wires versus magnetic field.

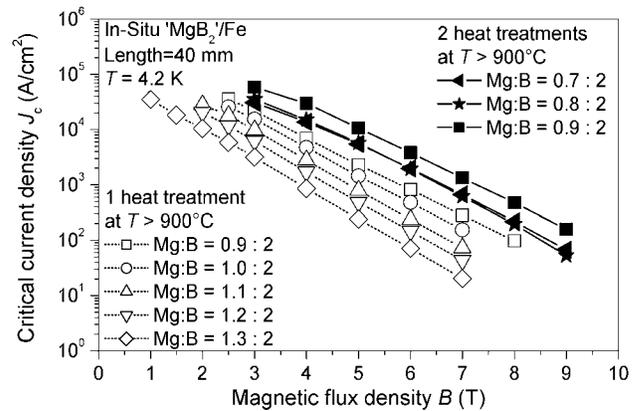

Fig. 7. Critical current density of once (open symbols) and twice (filled symbols) heat treated in-situ $MgB_2$/Fe wires with different Mg : B precursor ratio versus magnetic field.

The filament of the ex-situ wire looks much more homogeneous with only few boron rich grains (shown in Fig. 5) suggesting that a bigger part of the total cross section of the filament is capable to carry transport currents. However, we found, that the transport critical current density of the in-situ wire shown in Fig. 3 was even higher than the transport critical current density of the ex-situ wire. This is an indication for an improved quality of the percolation path in the in-situ wire.

### B. Critical Current Densities

Fig. 6 shows the critical current density $J_c$ of an as-swaged ex-situ $MgB_2$/Fe wire in comparison to a wire heat treated as shown in Fig. 1. In the magnetic field range of 4 - 8 T the critical current density of the heat treated wire is more than one order of magnitude higher than $J_c$ of the as-swaged wire. However, due to unsufficient thermal stabilization the heat treated wire burned at 4 T reaching $J_c = 1.57 \cdot 10^4$ A/cm$^2$. Due to the much smaller transport critical current density, the as-swaged wire could be measured in decreasing magnetic fields down to 0.25 T. The higher critical current density of the heat treated wire might be a result of the improvement of grain connectivity and a reduction of deformation induced stresses as reported by Suo et al. [5][5].

The critical current density $J_c(B)$ of in-situ wires with one final heat treatment at $T > 900°C$ and Mg : B ingot varying from 1.0 : 2 to 1.3 : 2 and of in-situ wires with two heat treatments at $T > 900°C$ and Mg : B varying from 0.7 : 2 to 0.9 : 2 is shown in Fig. 7. The critical current density of the wires with only one heat treatment above 900°C increases with decreasing Mg content. Strangely enough, $J_c$ of the wire with Mg : B = 0.9 : 2 exceeds $J_c$ of the wire with stoichiometric Mg : B ratio, although we expected higher current densities for wires with overstoichiometric Mg content. Because of Mg losses during the heat treatment the wires with stoichiometric or Mg poor Mg : B precursor ratio were supposed to contain large amounts of $Mg_{1-x}B_2$ or even $MgB_4$ or higher borides. However, the reason for the high transport critical current density of the wire with understoichiometric Mg : B precursor ratio can be found regarding the SEM picture and EDX analysis of this wire (Fig. 4). The matrix of the filament (light grey in the SEM picture) is Mg rich and is supposed to contain the superconducting phase. In contrast, the black grains contain mainly boron, only little magnesium difuses into the grains via the dendritic cracks within the grains. As the boron rich and therefore magnesium poor grains obey a very large part of the filament volume, the matrix of wires with stoichiometric or Mg rich precursor ratio is supposed to contain more Mg than that required for the $MgB_2$ formation.

In-situ wires with Mg : B ingot from 0.7 : 2 to 0.9 : 2, which were heat treated above 900°C twice, showed higher $J_c(B)$ values than the wires with Mg : B from 0.9 : 2 to 1.3 : 2 heat treated at these temperatures only once. Again, the critical current density of the wire with Mg : B = 0.9 : 2 was the highest, while wires with smaller Mg : B ratios showed smaller $J_c(B)$ values.

### C. Transition Temperature to Superconductivity

The real part of the ac-susceptibility $\chi'_{ac}$ of commercial Alfa Aesar powder and of ex-situ $MgB_2$/Fe wires made from this powder is displayed in Fig. 8. The transition to superconductivity of the commercial Alfa Aesar powder shows two phases with $T_{c1} = 38.5$ K and $T_{c1} = 37.4$ K (determined from the midpoint of the transition of each phase). The as-swaged $MgB_2$/Fe wire still shows a two-phase transition with unchanged onset temperature. However, the transition broadened strongly with the midpoint of the lower-temperature transition shifting down below 30 K. The strong broadening and temperature decrease of the transition to superconductivity might be a consequence of deformation-induced stresses which tend to reduce $T_c$ as shown in high-pressure experiments, where $T_c$ of $MgB_2$ powder was measured in a non-hydrostatic environment [10]. After a heat-treatment at temperatures above 900°C the transition to superconductivity of the ex-situ $MgB_2$/Fe wire is very narrow and the two phases cannot be distinguished anymore. The midpoint of the transition is $T_c = 38.1$ K with the onset being still unchanged. The upturn of $\chi'_{ac}$ below 37 K is of magnetic origin and supposed to be due to a reaction of the iron sheath with the filament since it was only observed in wires heat treated at temperatures above 900°C. In [5] the decrease of $T_c$ of $MgB_2$/Fe tapes after wire deformation is attributed to a weak link behavior and the increase of $T_c$ after annealing to an improvement of grain connectivity which would explain the much higher $J_c$ values of our heat treated $MgB_2$ wire. However, we do not think that this is the full story, as $T_c$ of the original Alfa Aesar powder is even higher than $T_c$ of the as-swaged wire, although the powder was filled loosely in the sample holder during $T_c$ measurement, i.e. with bad grain connectivity.

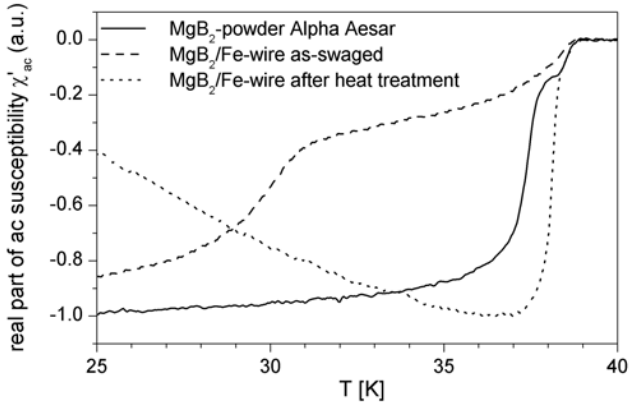

Fig. 8. Temperature dependence of the real part of the ac-susceptibility of $MgB_2$ powder and as-swaged and heat treated $MgB_2$/Fe wires. The upturn of the curves of the heat treated wire below $T_c$ is due to a magnetic backgound signal of the Fe sheath.

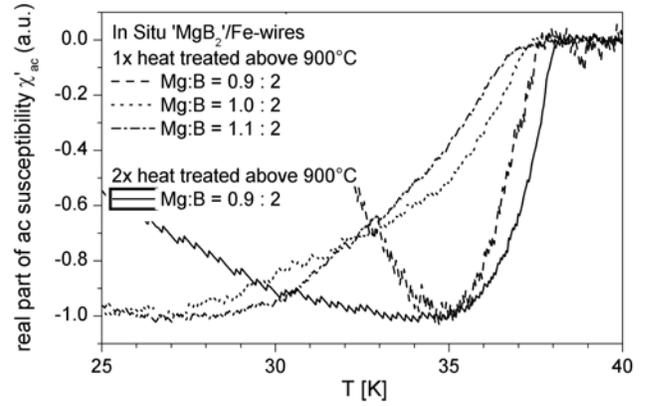

Fig. 9. Real part of the ac-susceptibility of in-situ $MgB_2$/Fe wires with different Mg : B precursor ratio. The upturn of the curves below $T_c$ is due to a magnetic backgound signal of the Fe sheath.

The real part of the ac-susceptibility $\chi'_{ac}$ of in-situ wires with Mg : B from 0.9 : 2 to 1.1 : 2 is shown in Fig. 9. With decreasing Mg : B ratio the transition of the wires which were heat treated above 900°C only once shift to higher temperatures. This result coinsides with results of Indenbom et al. [11] who showed $T_c$ of $Mg_{1-x}B_2$ to increase with increasing Mg deficiency. $T_c$ of the double heat treated wire with Mg : B = 0.9 : 2 was found to be higher than $T_c$ of the single heat treated wires, similar to the trend of the critical current densities $J_c(B)$. Although $T_c$ of this double heat treated wire is still approximately 1 K below $T_c$ of the heat treated ex-situ $MgB_2$-wire the critical current density $J_c(B)$ is slightly higher.

## IV. CONCLUSIONS

The critical current density and transition temperature to superconductivity of $MgB_2$/Fe wires strongly depend on the preparation process. We found that in-situ wires with an Mg : B precursor ratios of 0.9 : 2, which is slightly below the 'stoichiometric' ratio, showed the best $J_c(B)$ and $T_c$ values. $T_c$ and $J_c$ of ex-situ wires increase if the wires are heat treated above 900°C after the deformation process. Moreover, for in-situ wires an improved approach with two heat treatments is necessary to obtain comparable high $J_c$ and $T_c$ values. One $MgB_2$ formation heat treatment applied at 2/3 of the deformation process and the second as a final heat treatment.

Although the microstructure of ex-situ wires with commercial $MgB_2$ precursor seems to be much more homogeneous than that of in-situ wires prepared from elemental Mg and B, the critical current density of our best in-situ wire is even higher than $J_c$ of the best ex-situ wire. This is an indication of very good material quality in the percolation path of the filament. The still unsatisfying multi-phase microstructure of the filaments indicates much potential for further improvements of in-situ wires hopefully leading to improved superconducting properties.